\documentclass[a4paper]{article}

\usepackage{psfig,amsfonts,amssymb,amsthm}
\usepackage[totalwidth=17cm,totalheight=24cm]{geometry}

\newtheorem{prop}{Proposition}[section]
\newtheorem{lemma}[prop]{Lemma}

\newtheorem{thm}[prop]{Theorem}
\newtheorem{cor}[prop]{Corollary}

\newtheorem{df}[prop]{Definition}

\newcommand{\N}{{\mathbb N}}
\newcommand{\R}{{\mathbb R}}
\newcommand{\Z}{{\mathbb Z}}
\newcommand{\cH}{{\mathcal H}}

\newcommand{\Tr}{\rm{Tr}}

\newcommand{\Rh}{\rm{Rh}}

\newcommand{\D}{{\mathbb D}}
\newcommand{\eps}{\epsilon}

\newcommand{\Gd}{G^{\diamondsuit}}
\newcommand{\Gt}{\tilde{G}}

\newcommand{\Tt}{\tilde{T}}

\newcommand{\F}{F(\Tr(G), S^1)}
\newcommand{\Fpe}{F_{\geq}(\Tr(G), S^1)}
\newcommand{\Fsp}{F_{>}(\Tr(G), S^1)}

\begin{document}

\title{Rhombic embeddings of planar graphs with faces of degree $4$}

\author{Richard Kenyon\thanks{Laboratoire de Math{\'e}\-matiques, CNRS UMR
    8628, Universit{\'e} Paris-Sud, 91405 Orsay, France. 
\texttt{http://topo.math.u-psud.fr/\~{}kenyon}}
\and Jean-Marc Schlenker\thanks{
Laboratoire Emile Picard, UMR CNRS 5580,
Universit{\'e} Paul Sabatier,
118 route de Narbonne,
31062 Toulouse Cedex 4,
France.
\texttt{schlenker@picard.ups-tlse.fr;
  http://picard.ups-tlse.fr/\~{ }schlenker}. }}

\date{April 2003}

\maketitle

\begin{abstract}
Given a finite or infinite planar graph all of whose faces have degree $4$, 
we study embeddings in the plane in which all edges have length $1$,
that is, in which every face is a rhombus. We give a necessary and sufficient 
condition for the existence of such an embedding, as well as a description
of the set of all such embeddings. 

\bigskip

\begin{center} {\bf R{\'e}sum{\'e}} \end{center}
Etant donn{\'e} un graphe planaire, fini ou infini, dont toutes les faces
sont de degr{\'e} $4$, on {\'e}tudie ses plongements dans le plan dont toutes
les ar{\^e}tes sont de longueur $1$, c'est {\`a} dire dont toutes les faces sont
des losanges. On donne une condition n{\'e}cessaire et suffisante pour
l'existence d'un tel plongement, et on d{\'e}crit l'ensemble de ces
plongements. 
\end{abstract}

\bigskip

\section{Introduction}
\begin{df}
Given a planar graph $G$ all of whose faces
(except possibly for the outer face if $G$ is finite) 
have degree $4$,
a {\bf rhombic embedding} is an embedding of $G$ in $\R^2$ with the property 
that all edges are line segments and have length
$1$ (and hence each bounded face is a rhombus). 
\end{df}

Such embedding arise in  discrete complex analysis \cite{Duffin,Mercat}
and in statistical mechanics \cite{Kenyon, Mercat-thesis, Kenyon.ictp}. 
Here we study the spaces of such embeddings.

Our main results are the following.
\begin{enumerate}
\item We give a simple necessary and sufficient condition (Theorem \ref{planar})
for a planar graph to have a rhombic embedding in $\R^2$. 
\item We show that the 
space of rhombic embeddings of an infinite graph (with no unbounded faces) 
is a convex set when
parametrized by the rhombus angles.
There is a simple description of the extreme points of the closure of
this convex set. 
(Section \ref{convex} and Theorem \ref{extremepts}.)
\item We consider the space of periodic rhombic embeddings of a periodic planar
graph. It is the interior of a convex polyhedron. The area of the
fundamental domain provides a 
strictly convex functional on this polyhedron; we give a geometric
description of the unique critical point (Theorem \ref{maxarea}).
\end{enumerate}

\section{Background and motivations}
\subsection{Discrete complex analysis and statistical mechanics}
Rhombic embeddings first appeared in Duffin \cite{Duffin}, in the context
of discrete complex analysis. Indeed, as Duffin shows, there is a natural
way to define discrete analytic functions on graphs with rhombic embeddings,
which does not appear to generalize to arbitrary embeddings.  

Duffin's ideas were rediscovered by Mercat \cite{Mercat-thesis}
who used them to build up an extended theory of discrete holomorphy in 
one complex dimension. 

In \cite{Kenyon}, the closely related concept of
{\bf isoradial embeddings} (see the next section) was shown to be useful in the study of the 
so-called dimer model of statistical mechanics. By choosing 
edge interactions to be a particular function of
the corresponding edge lengths 
in the isoradial embedding, certain simplifications (commutation relations) appear which
do not appear for more general energies. These allow one to give explicit
solutions to correlation functions in the dimer model on isoradial graphs. 

Other statistical mechanical models such as the Ising model and more generally
the random cluster model also become simpler on isoradial graphs
\cite{Mercat-thesis, Kenyon.ictp}. In fact, even the simple random walk
behaves nicely, as is evidenced by an explicit formula for the Green's
function \cite{Kenyon}.

\subsection{Isoradial embeddings}
An {\bf isoradial embedding} (see \cite{Kenyon}) of a planar graph $G$
is a locally finite embedding in $\R^2$
with the property that each 
bounded face $f$ is a cyclic polygon (inscribable in a circle)
with circumcircle of radius $1$.
The center of the circumcircle $C$ is also called the center of $f$. 
The embedding is said to be {\bf convex} if the center of a face is
contained in the closure of the face. It is {\bf strictly convex}
if the center is in the interior of the face.

Given a planar graph $G$, the {\bf diamond graph} $\Gd$ associated to $G$ is
the graph 
whose vertices are the union of the vertices and faces of $G$, and with an
edge  
between each face and vertex on that face. The faces of $\Gd$ are the edges of
$G$. 
The faces of $\Gd$ are of degree $4$.

It is clear that a strictly convex isoradial embedding of a graph $G$
gives a rhombic embedding of $\Gd$. Conversely, a rhombic  embedding of $\Gd$ 
defines a strictly convex isoradial embedding of $G$. Furthermore every
planar graph with faces of degree $4$ arises as the diamond graph of another
planar graph. 
So the study of strictly convex isoradial embeddings and rhombic embeddings is
equivalent.

\section{Rhombic embeddings}
In this paper a planar graph will mean a graph with a preferred isotopy
class of embeddings 
in the plane. An embedding of such a graph will always mean an embedding in the
same isotopy class.

We will only deal with either finite graphs, all of whose faces except the
outer face 
have degree $4$,  or with infinite graphs with no unbounded
faces, that is, all of whose faces have degree $4$. In either case
we refer to such graphs as {\bf graphs with faces of degree $4$}, 
with the understanding
that we exclude the outer face from this restriction in the case of a finite
graph.  
Although the results in this section apply in greater generality,
for simplicity we deal only with these cases.

In a planar graph $G$ with faces of degree four, a  {\bf train track} is
a path of faces 
(each face being adjacent along an edge to the previous face)
which does not turn: on entering a face it exits across the opposite edge.
We assume that the train tracks extend 
in both directions as far as possible, that is, they are not embedded in
a longer train track. This means they are either periodic, or extend
infinitely far in both directions, 
or extend until they enter the outer face of $G$ if $G$ is finite.
Let $\Tr(G)$ denote the set of train tracks of $G$.
In a rhombic embedding, each rhombus in a train track has an edge
parallel to a fixed  
unit vector $u$.
In an oriented train track we choose the direction of $u$ so that when
the train runs down 
the track $u$ points from the right to the left. The vector $u$ is called the
{\bf transversal} of the oriented train track.

\begin{thm}\label{planar}
A planar graph $G$ with faces of degree $4$
has a rhombic embedding in the plane
if and only if the following two conditions
are satisfied
\begin{enumerate}
\item No train track path crosses itself or is periodic.
\item Two distinct train tracks cross each other at most once.
\end{enumerate}
\end{thm}

The conditions are clearly necessary: 
in a rhombic embedding each train track is a monotone path 
(in the direction perpendicular to its common parallel). 
Therefore a train track cannot cross itself. If two oriented train tracks with 
transversals $u$ and $v$ cross (with the first crossing from right to
the left of the second), 
the rhombus on which they cross has edges $u$ and $v$ and $u\wedge v>0$;
if they crossed 
again the first would cross from the left to the right of the second and
the rhombus would have the same two edges but in the reverse orientation,
that is, it would have negative area. 

It remains to construct a rhombic embedding for any graph satisfying the
conditions. Enumerate the distinct train tracks $\{t_1,t_2,\dots\}$. 
Let $G_k$ be the union of the faces
contained in the first $k$ train tracks $\{t_1,\dots,t_k\}$.
The $G_k$ are subgraphs of $G$ and $G=\cup_{k=1}^\infty G_k$.
Similarly let $G_k'$ be the subgraph of the dual $G'$ which is the 
union of the edge paths comprising the first $k$ train tracks.

\begin{lemma} \label{lm:plongement}
There exists a topological embedding $\rho$ of $G'$ in the unit disk $\D$, 
such that the image of each path of $G'$ (which corresponds to
a train track in $G$) 
is a smooth path connecting distinct
boundary points of the disk. 
\end{lemma}

\begin{proof}
Since $G'$ is planar, first choose a topological embedding of $G'$ in the disk.
For each $k$, by removing all but the first 
$k$ train tracks from this embedding,
we get topological embeddings of $G'_1,G'_2,\dots$,
in which the embedding of $G'_k$ is an extension of that of $G'_{k-1}$.

To make a sequence of smooth embeddings, with appropriate 
boundary behavior of the train tracks, proceed inductively as follows.
Choose any nontrivial chord for $G'_1$.
Suppose now that the smooth embedding of $G'_k$ has been defined, isotopic
to the original topological embedding. 
The graph $G'_{k+1}$ is obtained from $G'_k$ by adding a single
line $t_{k+1}$ 
which has a finite number of intersections with the lines present
in $G'_k$. On top of the smooth embedding of $G'_k$ draw 
in the train track $t_{k+1}$ as a smooth path, 
respecting its intersections (and their order) with each
of the $t_j$ for $j\leq k$, that is, so that the resulting embedding of 
$G'_k\cup t_{k+1}=G'_{k+1}$ is topologically
equivalent to the original topological embedding of $G'_{k+1}$. 
Since $t_{k+1}$ has only a finite number of intersections with $G'_k$,
it can be chosen so that its 
endpoints are distinct points on the boundary of the disk.
This defines the smooth embedding of $G'_{k+1}$. 
The union of these embeddings over all $k$ is a planar embedding of $G'$:
the image of any finite piece of $G'$ is fixed after a finite number 
of steps of this algorithm.
\end{proof}

To each oriented train track $t_j$ we associate the unit vector 
$u_j$ which is perpendicular
to the chord joining its two endpoints in the above embedding.
We construct the rhombic embedding of $G$ as follows. 
Each face of $G$ is crossed
by exactly two train tracks $t_j$ and $t_k$. To this face associate the rhombus
with edges $u_j$ and $u_k$.
Glue two rhombi together along an edge if they are adjacent faces in
$G$; this defines 
a simply connected locally Euclidean surface. It remains to show that
it is isometric to a subset of the plane. 

This surface has a natural projection $\phi$
to the plane, which clearly is locally
injective. We have to prove that $\phi$ is injective. 

\begin{lemma} \label{lm:translation}
Let $v,w$ be two vertices of $G$, and let $t_{k_1}, \cdots, t_{k_n}$ be the
train-tracks separating $v$ and $w$; assume these tracks are oriented
so that $v$ is on the right and $w$ is on the left. Then:
\begin{equation} \label{eq:sum} 
\phi(w)-\phi(v)=\sum_{j=1}^n u_{k_j}~. 
\end{equation}
\end{lemma}

\begin{proof}
Let $\Gamma$ be any path going from $v$ to $w$ in the 1-skeleton of $G$. 
$\Gamma$ is a sequence of edges of $G$, each corresponding to a train
track. Each of the train tracks $t_{k_1}, \cdots, t_{k_n}$ appears an odd
number of times, and is crossed once more from right to left than from left to
right. All the other train tracks are crossed the same number of times in both
directions. So the total translation vector of $\Gamma$ is a sum of terms
which pairwise cancel, except those in (\ref{eq:sum}). 
\end{proof}

Let $X$ and $Y$ be two distinct faces of $G$. We will show that their 
$\phi$-images have disjoint interiors.  
$X$ and $Y$ correspond to vertices $x$ and $y$, respectively, in the
graph $G'$. We choose $k$ so large 
that the train tracks containing $X$ and $Y$, as
well as the train-tracks which go between $X$ and $Y$, are in $\{ t_1, \cdots,
t_k\}$. 

\begin{lemma} \label{lm:path}
There exists an embedded path $\gamma:[0,1]\rightarrow \D^2$, with endpoints
on the boundary of the disk, which goes through $\rho(x)$ and $\rho(y)$, and
crosses each $\rho(t_j), 1\leq j\leq k$, at most once. 
\end{lemma}

\begin{proof}
Consider the restriction of the embedding $\rho$ to $G'_k$. 
We continuously deform it so that the path corresponding to  
each $t_j$ becomes a chord of the circle with the same endpoints.
Under this
deformation some topological changes may take place: a path may move past an
intersection of two other paths as in Figure \ref{reidemeister3}. 
\begin{figure}[htbp]
\centerline{\psfig{figure=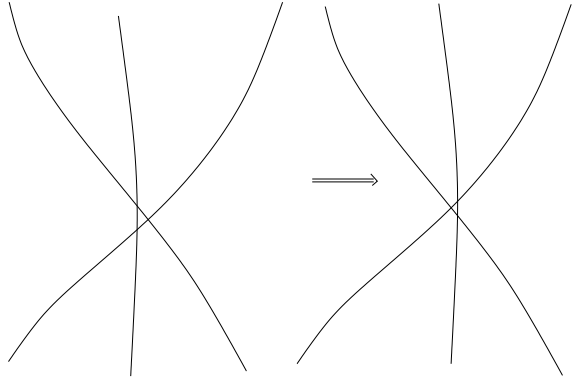,height=5cm}}
\caption{\label{reidemeister3} Moving a strand past an intersection. }
\end{figure}
As a result the deformed graph $\tilde G_k'$ is not 
necessarily isomorphic to $G_k'$, but can be obtained from $G_k'$ by
a finite sequence of these triple-crossing moves. Moreover there is a natural
bijection 
between the vertices of $\tilde G_k'$ and those of $G_k'$. In $\tilde G_k'$
draw $\gamma$ as a chord passing through $\rho(x)$ and $\rho(y)$.  
By a general position argument we can assume that no three chords meet
at a point, and $\gamma$ does not meet any other chord intersections.
Now undo the deformation, passing from $\tilde G_k'\cup\gamma$ to $G_k'$,
undoing the triple-crossing moves in sequence.
It is clear how to deform $\gamma$ at the same time,
keeping the property of $\gamma$ passing through $\rho(x)$ and $\rho(y)$, 
so that $\gamma$ never crosses any path more than once:
see Figure \ref{gammamove} which shows how $\gamma$ may be
deformed at each triple crossing. This completes the construction of $\gamma$.

\begin{figure}[htbp]
\centerline{\psfig{figure=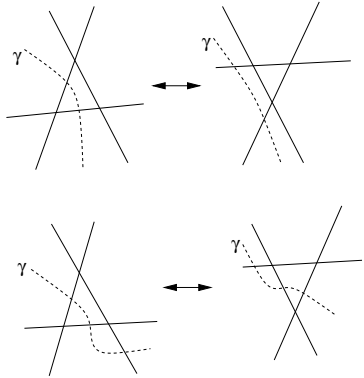,height=5cm}}
\caption{\label{gammamove} Moving $\gamma$ at a triple intersection.}
\end{figure}
\end{proof}

The fact that $X$ and $Y$ have disjoint images, and therefore the 
proof of Theorem \ref{planar}, follows from the next lemma. 

\begin{lemma}
Let $v$ and $w$ be vertices of $X$ and $Y$, respectively. The orthogonal
projection of $\phi(w)-\phi(v)$ on the oriented line $(\gamma(0)\gamma(1))$
has positive orientation.
\end{lemma}

\begin{proof}
Let $t_{k_1}, \cdots, t_{k_n}$ be the oriented train tracks which have $v$ on
their right and $w$ on their left. By the choice of $k$ above, $t_{k_j}\leq k$
for all $j\in \{ 1, \cdots, n\}$. By lemma \ref{lm:translation}:
$$ \phi(w)-\phi(v) = \sum_{j=1}^n u_{k_j}~. $$
But, for all $j\in \{ 1, \cdots, n\}$, $t_{k_j}$ intersects $\gamma$, so it 
has $\gamma(0)$ on its right
and $\gamma(1)$ on its left. By construction of $u_{k_j}$, it implies that
the orthogonal projection on $(\gamma(0)\gamma(1))$ of $u_{k_j}$ is positively
oriented, and the result follows. 
\end{proof}

\section{The space of rhombic embeddings}
In this section we suppose that $G$ is infinite.

\subsection{Linear parametrization}\label{convex}
A {\bf wedge} in a planar graph is a pair consisting of a face and a
vertex on that face. 
We let $W$ denote the set of wedges of a graph $G$.

A rhombic embedding determines an angle in $(0,\pi)$ 
for each wedge of $G$, satisfying some simple linear conditions. The
converse is also true.

\begin{thm} \label{embedding}
Let $\alpha:W\rightarrow (0, \pi)$ be a map such that:
\begin{itemize}
\item adjacent wedges on the same face have angles summing to $\pi$.
\item the sum of the angles around each vertex is $2\pi$.
\end{itemize}
Then $\alpha$ is obtained on a unique rhombic embedding $\phi$ of $G$ in
$\R^2$, up to an isometry. 
The image of $\phi$ is either the plane, a half-plane, or an
infinite strip.
\end{thm}

\begin{figure}[htbp]
\centerline{\psfig{figure=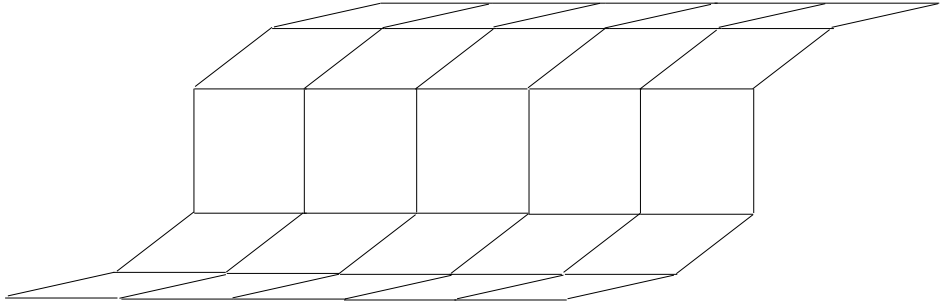,height=4cm}}
\caption{\label{strip-fig} An embedding of $\Z^2$ with image a strip.}
\end{figure}

Note that $\phi(G)$ can indeed be a half-plane or a strip: it is easy to
find examples of this when $G=\Z^2$ (see Figure \ref{strip-fig}). 
The proof of this
theorem will be given after the next lemmas.
\medskip

\begin{lemma}
Let $G_0$ be a finite graph with faces of degree $4$. Let $v_0$ be a
vertex of $G_0$ which is at (combinatorial) distance $k$ from the outer
face. For each $\epsilon>0$, there exists $\alpha>0$ such that, if $\phi$
is a rhombic immersion of $G$ in $\R^2$ and each face has area at most
$\alpha$, then:
\begin{itemize}
\item $\phi(G_0)$ is $\epsilon$-close to a line $L$ containing $\phi(v_0)$.
\item there exists a segment of $L$ of
length $2k$, centered at $\phi(v_0)$, which is $\epsilon$-close to
$\phi(G)$.  
\end{itemize}
\end{lemma}

\begin{proof}
$\phi$ is uniquely determined (up to translation) by the transverse
directions of the train-tracks in $G_0$. Moreover, if two intersecting
train-tracks have transverse directions which are neither close nor
almost opposite, the area of the faces can not all be small. Therefore, the
transverse directions of all train-tracks in $G_0$ are close to either
$u$ or to $-u$, for some $u\in S^1$. 

This already shows that $\phi(G_0)$ remains close to the line containing
$\phi(x_0)$ and of direction $u$. In addition, if $v$ is a
vertex of $G_0$ which is not adjacent to the outer face, then the edges
starting from $v$ can not be all pointing in directions close to $u$ 
or all in the directions close to $-u$ -- there must be some pointing
in both directions. 
Since $v_0$ is at distance at least $k$
from the outer face, the segment of length $2k$ centered at $\phi(v_0)$
and directed by $u$ is close to $\phi(G_0)$.
\end{proof}

\begin{lemma} \label{strip}
Let $(\phi_t)_{t\in
[0,1]}$ be a 1-parameter family of rhombic immersions of $G$ (i.e. not
necessarily globally injective). Suppose that, for all $t\in [0,1)$,
$\phi_t$ is an embedding. Then $\phi_1$ is an embedding, and 
$\phi_1(G)$ is either a plane, a half-plane or a strip. 
\end{lemma}

\begin{proof}
Define a {\bf boundary point} of $\phi_1$ as a point $x\in \R^2$ such
that any neighborhood of $x$ intersects an infinite set of rhombi of
$\phi(G)$. If $\phi_1$ has no boundary point, then it is a local
homeomorphism and it is proper, so that it is a global homeomorphism;
thus it is an embedding, with image $\R^2$. So we suppose that
$\phi_1$ has a boundary point $x_0$. 

Since the $\phi_t$ are embeddings for $t<1$, 
for any $k\in \N$, the sum of the areas of the
rhombi of $\phi_1(G)$ which are within distance at most $k$ from
$x_0$ is finite (bounded by the area of the disk of radius $k+1$). However,
by definition of a boundary point, there is a sequence of
vertices $v_n$ of $G$, $n\in \N$ such that
$(\phi_1(v_n))\rightarrow x_0$ and that the disks $D_n$ of
(combinatorial) 
radius $k$ centered on the $v_n$ are disjoint in $G$. Then the area of
the $\phi_1(D_n)$ go to $0$ as $n\rightarrow \infty$. 

By the lemma above, for $n$ large enough, the $\phi_1(D_n)$ are
each close to a line going through $x_0$. By construction of $t_0$, they
are disjoint, so they are all close to a fixed line $L\ni
x_0$. Moreover, still by the lemma, there is a segment $\sigma$ of $L$ of
length $2k$ centered at $x_0$ which is arbitrarily close to the
$\phi_1(D_n)$ for $n$ large enough. 

Therefore, all points of $\sigma$ are boundary points of
$\phi_1$. This is true for all $k\in \N$, so the line $L$ is made up completely
of boundary points of $\phi_1$. What is more,
$\phi_1(G)$ can not be on both sides of
$L$, since otherwise $G$ could not be connected. Since the same can be
said of all the boundary points of $\phi_1$, we have that
$\phi_1(G)$ is bounded by
disjoint lines. There can be either one or two of those lines; in the
first case $\phi_1(G)$ is a half-plane, in the second case it is a strip. 
This description shows that $\phi_1$ is globally injective, so it is an
embedding. 
\end{proof}

It is now possible to prove Theorem \ref{embedding}.

\begin{proof} 
Given a set of angles satisfying these two linear conditions, 
there corresponds at most one rhombic embedding up to global isometry.
Conversely, it is clear that such a choice of angles determines a
locally injective map from $G$ to $\R^2$, but it remains to show that it
is globally injective.

The space of maps $\alpha:W\rightarrow (0,\pi)$ satisfying the two
linear conditions is convex. We know from Theorem \ref{planar}
that there exists a choice of angles, say $\alpha_0$, giving an
embedded graph. We call $(\alpha_t)_{t\in [0,1]}$ the affine 
parametrized segment of maps $\alpha_t:W\rightarrow (0,\pi)$ between
$\alpha_0$ and $\alpha_1=\alpha$, and $(\phi_t)_{t\in [0,1]}$ the
associated 1-parameter family of rhombic immersions of $G$. 

We will show that $\phi_t$ is an embedding --- i.e. is globally
injective --- for all $t\in [0,1)$. Lemma \ref{strip} then shows that
$\phi_1$ is an embedding, with image a plane, half-plane or strip. 
Suppose $\phi_t$ is not an embedding for all $t$; let
$t_0$ be the infimum of the $t\in [0,1]$ such that $\phi_t$ is not
globally injective, and we suppose that $t_0<1$.  

Lemma \ref{strip}, applied to the 1-parameter family $(\phi_t)_{t\in
  [0,t_0]}$, shows that $\phi_{t_0}$ is an embedding, and that its image
is either a plane, a half-plane or a strip. 
Moreover, the definition of $(\phi_t)$ from the affine segment
$(\alpha_t)_{t\in [0,1]}$ shows that no rhombus angle goes to $0$ as
$t\rightarrow t_0$. Therefore $\phi_{t_0}$ can not have boundary
points, and its image is a plane. For $t\geq
t_0$ close enough to $t_0$, each of the rhombus angles remains bounded
from below, so $\phi_t$ is still proper, and
thus an embedding. This contradicts the definition of $t_0$. So $\phi_t$
is an embedding for all $t\in [0,1]$. 
\end{proof}

\subsection{Convexity}
Theorem \ref{embedding} proves that
the space of rhombic embeddings is parametrized by the choice
of an angle for each wedge, satisfying the above linear conditions.
Thus the space of rhombic embeddings is the intersection of a cube
$(0,\pi)^W$ with the set of linear subspaces defined by these
constraints. Thus it is a convex set, which we denote $\Rh(G)$.

\medskip

Given a graph $G$ with faces of degree $4$, 
we will call $\Tr_+(G)$ the set of oriented
train-tracks. So $\Tr_+(G)$ has two elements for each train track $t\in
\Tr(G)$, one for each choice of orientation of $t$. For each $t\in
\Tr_+(G)$, we call $-t$ the oriented train-track which is the same as
$t$ but with the opposite orientation. 

We consider now only graphs which satisfy the hypothesis of Theorem
\ref{planar}. In particular, two oriented 
train-tracks $t,t'$ with $t\neq \pm t'$ intersect at most
once.  Given two distinct 
oriented train-tracks $t, t'\in \Tr_+(G)$, with $t'\neq\pm t$, we will
say that $t'$ has {\bf positive intersection} with $t$ if $t$ and $t'$
intersect, and the intersection has positive orientation, that is,
if $t'$ crosses $t$ from right to left. 

Let $\cal D$ be the directed graph whose vertices are the 
oriented train tracks, with an edge going from $t$ to $t'$ if and only if
$t'$ intersects $t$ positively. In particular, if $t'$ intersects $t$
positively, then $\cal D$ has a cycle $t\to t'\to -t\to -t'\to t$. 

Let $\phi\in \Rh(G)$ be a rhombic embedding of $G$. Then each oriented
train-track has a transversal, which is a unit vector, i.e. an element
of $S^1$. Thus $\phi$ induces a map $\Tr_+(G)\rightarrow S^1$, with the
property that $\phi(-t)=\phi(t)+\pi$. We will see below that
this map is strictly order-preserving in the following sense.

\begin{df} \label{df:pos-inter}
\begin{enumerate}
\item We call $\F$ the space of maps $\theta$ from $\Tr_+(G)$ to $S^1$ such
that, for all $t\in \Tr_+(G)$, $\theta(-t)=\theta(t)+\pi$.
\item A map $\theta\in \F$ is {\bf 
  strictly order preserving} if, for each $t, t'\in \Tr_+(G)$ such that
$t'$ has positive intersection with $t$,
$\theta(t')\in (\theta(t), \theta(t)+\pi)$. We call $\Fsp$ the space of
strictly order-preserving elements of $\F$. 
\item A map $\theta\in \F$ is {\bf order-preserving} if:
\begin{itemize}
\item 
For each $t, t'\in \Tr_+(G)$ such that $t'$ intersects $t$ positively,
$\theta(t')\in [\theta(t), \theta(t)+\pi]$.
\item 
Each subgraph of $\cal D$ consisting of train tracks having a fixed
$\theta$-value is acyclic (has no oriented cycles). 
 \end{itemize}
We call $\Fpe$
the space of {\bf order-preserving elements} of $\F$.
\end{enumerate}
\end{df}

\begin{lemma} \label{lm:rhombic}
A map $\theta\in \F$ determines a rhombic embedding of
$G$ if and only if it is strictly order-preserving. 
\end{lemma}

\begin{proof}
Given a rhombic embedding of $\phi$ of $G$, we have already mentioned
that it induces a map $u\in \F$. If two oriented train-tracks $t, t'\in
\Tr_+(G)$ intersect, then the intersection rhombus has positive
orientation and this means that, if $t'$ has positive intersection with
$t$, then $u(t')\in (u(t), u(t)+\pi)$. Thus $u$ is strictly
order-preserving. 

Conversely, let $u\in \Fsp$. Let $w=(f,v)$ be a wedge in $G$, where $f$
is a face and $v$ is a vertex of $f$. Let $t_1$ and $t_2$ be the two
oriented train-tracks such that the oriented edges $e_1$ and $e_2$ of
$f$ starting from $v$ are the oriented transverse directions of $t_1$
and $t_2$, respectively. Define a map $\alpha:W\rightarrow \R$ by
setting $\alpha(w)$ equal to the oriented angle between $u(t_1)$ and
$u(t_2)$. Since $u$ is strictly order-preserving, $\alpha$ takes its
values in $(0,\pi)$. It is clear that $\alpha$ satisfies the two linear
conditions in Theorem \ref{embedding}.

Thus Theorem \ref{embedding} shows that $\alpha:W\rightarrow (0,\pi)$ is
obtained on a rhombic embedding of $G$, for which --- up to an isometry
in $\R^2$ --- the transverse direction of each oriented train-track $t$
is given by $u(t)$. 
\end{proof}

Clearly two maps $\theta, \theta'\in \Fsp$ determine the same rhombic
embedding, up to a global isometry, if and only if there exists a
constant $\theta_0$ such that, for all $t\in \Tr_+(G)$,
$\theta'(t)=\theta(t)+\theta_0$. Therefore, $\Rh(G)$ is the quotient of
$\Fsp$ by $S^1$. 

\begin{lemma} \label{lm:closure}
The closure of $\Fsp$ is $\Fpe$. 
\end{lemma}

\begin{proof}
Let $\theta\in\F$. We want to show that the following statements are
equivalent: 
\begin{enumerate}
\item $\theta$ is order-preserving.
\item for all $\epsilon>0$, there exists $\theta'\in \Fsp$ which is
  strictly order-preserving and such that, for all $t\in 
  \Tr_+(G)$, $|\theta'(t)-\theta(t)|< \epsilon$. 
\end{enumerate}

It is clear that (2) implies (1), so we only have to prove the
converse. Let $\theta\in \F$ be
order-preserving. 

We can partition the set of train tracks into sets which are the maximal
components of $\cal D$ having the same $\theta$-value.
Without loss of generality we assume that the value of $\theta$ is
different on each component. 
Enumerate the components $C_1,C_2,\dots$; we can further assume by perturbing
$\theta$ slightly (keeping the value constant on each $C_i$) that
for $i<j$ the $\theta$ values of $C_i$ and $C_j$ are at least $\eps/2^j$ apart.

We now perturb the $\theta$-values within each $C_i$ so that the new values
lie in $(\theta-\eps/2^{i+1},\theta+\eps/2^{i+1})$. But $\cal D$
restricted to $C_i$ 
is an acyclic connected graph; as such it represents a partial order of its 
(at most countably many) vertices.
Assign values in $(\theta-\eps/2^i,\theta+\eps/2^i)$ to the train tracks
in a way which is compatible with this (strict) partial order. 
This completes the construction.
\end{proof}

\begin{lemma} \label{lm:extension}
The natural map from $\Fsp$ to $\Rh(G)$ extends continously to a map
from $\Fpe$ to the closure of $\Rh(G)$ (for its affine structure). The
induced map from $\Fpe/S^1$ to $\overline{\Rh(G)}$ is one-to-one.
\end{lemma}

\begin{proof}
It is clear that the map from $\Fsp$ to $\Rh(G)$ extends continously to
a map from $\Fpe$ to $\overline{\Rh(G)}$, which is surjective by
construction.

Given $u\in \Fsp$, $u$ is uniquely determined --- up to an isometry in
$\R^2$ --- by the angles at the wedges, which are now in $[0,\pi]$,
i.e. by the image of $u$ in $\overline{\Rh(G)}$.
So the induced map from $\Fpe/S^1$ to $\overline{\Rh(G)}$ is injective.
\end{proof}

As a consequence, the affine structure on $\Rh(G)$ can be obtained from
the parametrization by the wedge angles, but also
from the affine structure on $\Fsp$ through the quotient by
$S^1$. This second description will be useful below
to understand the extreme points of this convex set. 

\subsection{The group of circle homeomorphisms}
A consequence of the definitions of $\F, \Fpe$ and $\Fsp$ is that, for
each $G$, 
there is a canonical action on each of those convex sets of the group of
homeomorphisms of the circle.

\begin{df} We call $\cH$ the group of homeomorphisms
  $\phi:S^1\rightarrow S^1$ such that, for all $x\in S^1$,
  $\phi(-x)=-\phi(x)$.
\end{df}

There is a natural action of $\cH$ on $\F$, defined as follows. Let
$\phi\in \cH$. For 
each $\theta\in \F$ and each $t\in \Tr_+(G)$, let 
$\phi^*(\theta)(t):=\phi(\theta(t))$. It is clear that this defines an
action of $\cH$ on $\F$. Moreover, given $\phi\in \cH$, it is also clear
that an element 
$\theta\in \F$ is (strictly) order-preserving if and only if
$\phi(\theta)$ is (strictly) order-preserving, and therefore $\cH$ also
acts on $\Fpe$ and $\Fsp$. Note that the action of $\phi$ on $\F$ is not
affine. 

By lemmas \ref{lm:rhombic} and \ref{lm:closure}, $\Fpe$ is a closed convex
set. Thus it has a natural stratifications, with one codimension $0$
cell corresponding to $\Fsp$, and higher-dimensional stratums
corresponding to some intersecting train-tracks having the same image in
$S^1$. 

\begin{lemma} \label{lm:stratification}
The action of $\cH$ preserves each stratum of $\Fpe$. 
\end{lemma}

\begin{proof}
This is again a consequence of the definitions. 
\end{proof}

\subsection{Vertices}

\begin{thm} \label{extremepts}
The extreme points of $\Rh(G)$ are in one-to-one correspondence with
the maps $\theta:\Tr_+(G)\rightarrow \{-1,1\}$ satisfying
$\theta(-t)=-\theta(t)$, such that the subgraphs ${\cal D}_1$ and 
${\cal D}_{-1}$ of $\cal D$, consisting of vertices with values $\pm 1$
respectively, are acyclic.
\end{thm}

This theorem is a direct consequence (through the quotient by $S^1$) of 
the following lemma on the geometry of $\Fpe$. To state it, we introduce
a simple notation. Let $T\subset \Tr_+(G)$ be a subset such that, for
each $t\in \Tr_+(G)$, either $t\in T$ or $-t\in T$, but not both
($T$ is a choice of orientation for each train track). Then,
for $\alpha\in S^1$, define:
$$ \begin{array}{cccl}
  \theta_{T, \alpha}: & \Tr_+(G) & \rightarrow & S^1 \\
  & t & \mapsto & \left\{
    \begin{array}{cc}
      \alpha & \mbox{ if} ~ t\in T \\
      \alpha+\pi & \mbox{if} ~ t\not \in T
    \end{array}
\right.
\end{array} $$
Clearly, for all $T$ and $\alpha$, $\theta_{T, \alpha}\in \F$. 

\begin{lemma} \label{lm:extreme}
If $\theta_{T, \alpha}\in \Fpe$, then the set of $\theta_{T, \beta}$ for
$\beta\in S^1$ is a $1$-dimensional stratum of $\Fpe$. Moreover, all
$1$-dimensional stratums of $\Fpe$ are of this form. 
\end{lemma}

\begin{proof}
The first point is obvious. 
Given $\theta\in \Fpe$, the orbit ${\cal H}\theta$ is
one-dimensional if and only if $\theta$ takes only two values,
$\beta$ and $\beta+\pi$ for some $\beta\in S^1$. Therefore the stratum is 
one-dimensional only in this case. This proves the second
point. 
\end{proof}

\subsection{Extreme points when $G$ is periodic}
In this subsection we consider a graph $G$ embedded in the torus (with
faces of degree $4$) and its lift $\Gt$ to a periodic graph in $\R^2$. 
The extreme points of the space of rhombic embeddings of $\Gt$ are
described by Theorem \ref{extremepts} above.
Associated to an extreme point $\theta_{T,0}$ is a 
map from the vertices of $\Gt$ to $\Z$ with the property
that the difference between the values at the endpoints of each edge
differ by either $1$ or $-1$: the difference of the map along an edge $e$
which is the transversal of train track $t\in T$ is  $1$ (and $-1$ if $t\not\in T$).

Let $\pm t_1,\dots,\pm t_k$ be the train tracks of $G$. Let ${\cal D}(G)$ be the finite
directed graph of positive intersections. 

\begin{thm}
The extreme points $T$ of 
$\Rh(\Gt)$ can be described as follows. 
Assign elements of $\{+,-,0\}$ to vertices of
${\cal D}(G)$ so that the subgraph consisting of $\{0,+\}$ has no directed
cycle  
and the subgraph consisting of $\{0,-\}$ has no directed cycle.
To each such assignment is associated a collection of extreme points of $\Gt$
where $T$ is the union of the set of train tracks projecting to $+$-train
tracks 
of ${\cal D}(G)$ and an arbitrary subset of the
 train tracks projecting to $0$-train tracks of ${\cal D}(G)$. Conversely any
 extreme 
point of $\Rh(\Gt)$ arises in this fashion. 
\end{thm}

\begin{proof}
The graph ${\cal D}(\Gt)$ is obtained by replacing each vertex of 
${\cal D}(G)$ by a copy of $\Z$ and each edge from $v$ to $v'$ by the
collection  
of all edges from the copy of $\Z$ at $v$ to that at $v'$. 
To a subset $T$ of  vertices of ${\cal D}(\Gt)$, assign elements of
$\{0,+,-\}$ to  
${\cal D}(G)$ as follows: if every vertex of ${\cal D}(\Gt)$
 over $v\in {\cal D}(G)$ is in $T$, assign 
$+$ to $v$. If no vertex is in $T$, assign $-$. In the remaining case
assign $0$. 
Then a ``$+$'' component of $ {\cal D}(\Gt)$ contains a cycle if and only if
a $\{0,+\}$-component of ${\cal D}(G)$ contains a cycle, and similarly  
a ``$-$'' component of $ {\cal D}(\Gt)$ contains a cycle if and only if
a $\{0,-\}$-component of ${\cal D}(G)$ contains a cycle.
\end{proof}
 
This result can be formulated differently in terms of the directions of
the train tracks with image $1$ or $-1$ in an extreme point of
$\Rh(\Gt)$. 

\begin{cor}
The extreme points of $\Rh(\Gt)$ are in one-to-one correspondence with
the couples $(T_0, \Tt_0)$, where $T_0$ is a maximal subset of
$\Tr_+(G)$ of non-intersecting train-tracks with the same orientation,
and $\Tt_0$ is an arbitrary non-empty subset of $\Tr_+(\Gt)$ of train-tracks
projecting to the elements of $T_0$. The correspondence is achieved by
sending $(T_0, \Tt_0)$ to the map $\theta:\Tr_+(\Gt)\rightarrow \{
-1,1\}$ such that $\theta(t)=1$ if and only if either $t\in \Tt_0$,
or $t$ has positive intersection with a train-track projecting to $T_0$. 
\end{cor}

\paragraph{Example: $\Z^2.$} 
The graph ${\cal D}(\Z^2)$ has 4 vertices, corresponding to 
$\pm t_h$, the 
left- and right-oriented horizontal train track and $\pm t_v$,
 the up- and down-oriented
vertical train track; and four edges, $t_h\to t_v\to-t_h\to-t_v\to t_h$.
In an extreme point $T$, either $t_h=\pm $ and $t_v$ is arbitrary,
or $t_v=\pm $ and $t_h$ is arbitrary. 

\section{Periodic rhombic embeddings}
\subsection{Existence}
For graphs on a torus we have the following analog of Theorem \ref{planar}.

\begin{thm} \label{periodic}
Suppose $G$ is a finite graph embedded on a torus, and each face has degree
$4$. 
Then $G$ has a rhombic embedding on a torus if and only if 
the following two conditions are satisfied:
\begin{enumerate}
\item Each train track is a simple closed curve.
\item The lift of two train tracks to the universal cover intersect at
  most once. 
\end{enumerate}
\end{thm}

\begin{proof} The conditions are clearly necessary.
The sufficiency is proved by construction: let $(p_i,q_i)$ be the integer
homology class 
of $t_i$. Define $u_i$ to be the unit vector perpendicular to $(p_i,q_i)$. 
To see that this defines an embedding, it suffices to show that each
rhombus has positive area. However if oriented train tracks $t_i$ and
$t_j$ intersect, 
with $t_i$ crossing $t_j$ from right to left, then
$p_iq_j-p_jq_i>0$ since it counts the algebraic intersection number.
This implies that $u_i\wedge u_j>0$. 
\end{proof}

\subsection{Asymptotic directions} \label{sub:asymptotic}
Consider a graph $G$ with faces of degree $4$ embedded on a torus,
satisfying the conditions of Theorem \ref{periodic}.
Each train track $t$ is a simple closed curve on the torus; the homology
class of this 
curve is denoted $[t]\in\Z^2$. 
Two train tracks of homology classes $(p_1,q_1)$ and $(p_2,q_2)$
intersect $|p_1q_2-p_2q_1|$ times.

Given a rhombic embedding of $G$ on a torus, the lift of an oriented
train track $t$ to the universal
cover $\R^2$ is a periodic curve and hence has an {\bf asymptotic
  direction} $a(t)\in S^1$. It also has an {\bf asymptotic vector} $v(t)$,
which is the difference in $\R^2$ between two consecutive lifts of a
point of $t$. 
The asymptotic vector of a train track $t$ is of course the sum of
the transverse vectors of all rhombi comprising $t$. That is, we have
$$v(t_i)=\sum_j b_{i,j}u_j,$$ 
where the sum is over
all train tracks, $u_j$ is the transversal of oriented train track $j$, 
and $b_{i,j}$ is the number of intersections of train track $j$ with
train track $i$: 
$b_{i,j}=p_iq_j-p_jq_i$. 
The asymptotic direction is the unit vector $a(t_i)=v(t_i)/\|v(t_i)\|$. 

\subsection{The space of periodic rhombic embeddings}
Let $\Rh(G)$ be the space of rhombic embeddings of $G$ on a torus 
(the flat metric on the torus may be a function
of the embedding).
The argument of section \ref{convex} can still be used, but we now have
to consider only the finite set of train-tracks on the torus.
The proof of Theorem \ref{planar} --- in a simplified form --- shows
that $\Rh(G)$ is convex when
parametrized by the rhombus angles, but it is now finite-dimensional, of
dimension $|\Tr(G)|-1$. 

The characterization of its extreme points can be done just as in
Theorem \ref{extremepts} and lemma \ref{lm:extreme} above. We can still
consider $\Rh(G)$ as the quotient of $\Fsp$ by the action of $S^1$, and
the argument using the action of $\cH$ still shows that the
$1$-dimensional stratums of $\Fpe$ correspond to its elements which take
only two values $\beta$ and $\beta+\pi$. 

In particular $\overline{\Rh(G)}$ is a convex polyhedron of dimension
$|\Tr(G)|-1$.

\subsection{Canonical embedding}
On a torus, there is a ``best" rhombic embedding of any given graph;
it has a simple geometric characterization but is also obtained as the
rhombic embedding maximizing the area. 

\begin{thm}\label{maxarea} 
On $\Rh(G)$ the area function of the torus is a strictly concave function 
of the rhombus angles. There is a unique point maximizing the area, it
is characterized by 
the fact that the transverse direction of each train track is orthogonal
to its asymptotic direction.
\end{thm}

\begin{proof}
The area of a rhombus of angle $\theta$ is $\sin\theta$,
which is a strictly concave function of $\theta\in(0,\pi)$. This implies
that the total area, which is the sum of the areas of the individual rhombi,
is concave as a function of the angles. 
To show strict concavity, it suffices to show 
strict concavity under any one-parameter perturbation. 
But any perturbation must change the angle of some rhombus, and the
contribution from the area of this rhombus is strictly concave. So the
area has at most one critical point on $\Rh(G)$. 

Let $R$ be a rhombic embedding of graph $G$. The area of $R$ is:
\begin{equation} \label{eq1}
A(R) =  \sum_{1\leq i\neq j\leq t} b_{i,j} u_i\wedge u_j = \sum_{1\leq i\neq j\leq
  t} b_{i,j} \sin(\theta_j-\theta_i)~, 
\end{equation}
where $t$ is the number of train-tracks in $G$, $u_i=e^{i\theta_i}$ and
the sum runs over all train-tracks in $G$, with an orientation chosen
for each.  
As a function of $u_i$ it is a constant plus $u_i\wedge (\sum_j
b_{i,j}u_j)=u_i\wedge a(t_i)$. 
In particular since $u_i$ is a unit vector this quantity is critical if
and only if, for each $i$, $u_i$ is orthogonal to $a(t_i)$. So, if the
area does have a critical point on $\Rh(G)$, it is as described in the
theorem. 

For each $1\leq i\leq t$, let $\alpha_i$ be the angle of the asymptotic
direction of $t_i$. We define a vector field $V$ on $\Rh(G)$ by:
$$ \forall i\in \{ 1, \cdots, t\}, d\theta_i(V) =
\cos(\alpha_i-\theta_i)~. $$
Clearly $V$ vanishes exactly at the critical points of the
area. In fact, $A$ is increasing along the integral curves of $V$:
$$ dA(V) = \sum_i du_i(V)\wedge \sum_j b_{i,j}u_j = \sum_i
(-\cos(\theta_i -\alpha_i) Ju_i) \wedge v(t_i)~, $$
$$ dA(V) = \sum _i
\cos(\theta_i-\alpha_i) \langle u_i, v(t_i)\rangle  = \sum_i \|
v(t_i) \|\cos^2(\theta_i- \alpha_i)\geq 0~. $$ 

To prove that $\Rh(G)$ contains a critical point of $V$, it is therefore
sufficient to prove that, at any point of $\partial \Rh(G)$ where the area is
positive, $V$ points towards the interior of $\Rh(G)$. Let
$\theta^0=(\theta^0_i)_{1\leq i\leq t}\in \partial \Rh(G)$; the face of
$\partial\Rh(G)$ containing $\theta^0$ is characterized by a set of equalities
of the form $\theta_{i_p}=\theta_{j_p}$ for different values of $p$,
where --- without loss of generality --- $t_{j_p}$ has positive
intersection with $t_{i_p}$. 

For each $1\leq i\leq t$, let $\alpha_i^0$ be the angle of the
asymptotic direction of $t_i$ for $\theta^0$.
Then $\alpha_{i_p}^0, \alpha_{j_p}^0\in (\theta_{i_p}^0-\pi,
\theta_{i_p}^0)$ 
and $\alpha_{j_p}^0>\alpha_{i_p}^0$ (because $t_j$ has positive
intersection with $t_i$ and $A>0$). Therefore
$\cos(\theta_{j_p}^0-\alpha_{j_p}^0) >
\cos(\theta_{i_p}^0-\alpha_{i_p}^0)$, and 
thus $V$ points towards the interior of $\Rh(G)$. 
\end{proof}

\begin{cor} \label{lm:orthogonal}
A rhombic graph on a torus has maximal area if and only if, for each
train track $t$, 
the transverse direction of $t$ is orthogonal to its asymptotic
direction $a(t)$.  
\end{cor}

\section{Parallelogram embeddings}
A generalization of rhombic embeddings is parallogram embeddings, wherein
each face is mapped to a parallelogram. Given a parallelogram embedding, 
a unique rhombic embedding can be obtained by replacing each edge with the unit
length vector in the same direction. 
Conversely, the set of parallelogram embeddings associated to a
given rhombic embedding is obtained simply by
replacing each edge in a train track
with a real multiple (a different real multiple for each train track).
Thus the combinatorial and topological behavior of parallelogram embeddings
can be understood from the underlying rhombic embedding.

\section*{Acknowledgements}

The second author would like to thank Gilbert Levitt for some
remarks related to this work.

\end{document}